\documentstyle[epsf,12pt]{article}
\begin{document}
\begin{center}
\Large{R.S. Longacre$^a$, S.J. Lindenbaum$^{a,b}$\\
$^a$Brookhaven National Laboratory, Upton, NY 11973, USA\\
$^b$City College of New York, NY  10031, USA\footnote{This research was 
supported by the U.S. Department of Energy under Contract No. 
DE-AC02-98CH10886 and the City College of New York Physics Department}}
\end{center}
 
\begin{abstract}
 Four separate experiments(Etkin et al.[1-4]), observing the OZI forbidden
disconnected reaction $\pi^- p \rightarrow \phi \phi n$ with increasing 
statistics were consistent. These experiments very selectively completely
broke down the OZI suppression by 3 $\phi \phi$ resonances with 
$I^GJ^{PC} = 0^+2^{++}$ in the observed mass region 2.038 to 2.600 GeV. The 
only viable proposed explanation has been that the $I^GJ^{PC} = 0^+2^{++}$ 
glueball expected in this mass region caused the hard glue in the 
disconnection to resonate and very selectively breakdown the OZI suppression 
for its quantum numbers only. Recently a p p central production spin analysis 
found the $f_2(1950)$ had a dominant decay mode $f_2(1270) \pi \pi$
(Barberis et al. \cite{4pi}). We consider if it is related to the $\phi \phi$ 
resonances, and find that it likely is. 

Keywords: glueball or glueballs, OZI, $2^{++}$ glueball searches,
$2^{++}$ glueball predictions by Lattice Gauge theory, LGT glueball
predictions.

E-mail address: longacre@bnl.gov (R.S. Longacre), lindenbaum@bnl.gov 
(S.J. Lindenbaum).

\end{abstract}

\section{Introduction}

 The first critical test of non-perturbative QCD was the prediction of
a spectrum of multi-gluon resonant states(glueballs)by Lattice Gauge
Theory(LGT)[6-9]. Even phenomenological work made the existence 
of glueballs inescapable.

Unfortunately two decades later there is no consensus on the existence of 
even one glueball. Obviously the glueball spectrum is accompanied by the
many other varieties of quark built states and also possibly hybrids.
To date LGT assumes pure glue and has not solved the problem of including
quarks capable of predicting both spectra, and their mixing. If a glueball
has the same quantum numbers as a quark state which it overlaps in mass, 
it is expected from quantum mechanics that they would mix, and the mixed quark 
built states would have substantial glue. Thus it became clear that a filter
which passes glueballs or glue rich quark-glue mixed states, and rejects 
quark built states was extremely desirable for discovering glueballs, and
the accompanying substantially mixed with glue states if they existed.
 
The strong interaction OZI disconnected diagram $\pi^- p \rightarrow \phi
\phi n$ was proposed and used[1-4]  as such a filter. The two hard 
gluons bridging the disconnection would by OZI suppression reject quark built
states. However in a mass region where a glueball with the quantum numbers 
contained in the $\phi \phi$ system occurred, the gluons would resonate 
to form the glueball, break down the OZI suppression and let the glueball 
and mixed quark plus glueball states pass through. The mass region accessible 
to the $\phi \phi$ experiment\cite{phiphi4} 2 to 2.6 GeV was ideal for 
detecting the $2^{++}$ glueball[6-9].

\section{The $\phi\phi$ Resonances and Discussion}

The OZI rule was found to be completely broken down, very selectively, in
the reaction $\pi^- p \rightarrow \phi \phi n$ , by three resonance states 
with quantum numbers $I^GJ^{PC} = 0^+2^{++}$. These are  the quantum numbers 
of the $2^{++}$ glueball expected[6-9] in this mass region. The $\phi\phi$ 
production t dependence was sharply peaked forward with the shape consistent 
with $\pi$ exchange. This implies that the disconnected light quark lines 
ended in a $\pi^+ \pi^-$ annihilation bridged by gluons to the strange quark 
lines of the $\phi\phi$ system. Since $\pi$ exchange would only allow isoscalar
glueballs ($J^{PC} = 0^{++}, 2^{++}, 4^{++}$). The states were originally 
listed as $g_t$,$g'_t$,$g''_t$(are now listed as 
$f_2(2010),f_2(2300),f_2(2340)$ states) in the PDG summary table as 
established. They were found to contain approximately 87\% of the events in 
the sample of the $\phi K^+ K^- n$ as $\phi \phi$ resonances.
The $\phi K^+ K^- n$ reaction is OZI allowed, while the $\phi \phi n$ reaction
violates OZI and is 7.7 times bigger than the allowed reaction. The 13\%
$\phi K^+ K^- n$ events contain 8.6\% incoherent background without structure,
plus 4\% $I^GJ^{PC} = 1^+1^{--}$, and 0.4\% $I^GJ^{PC} = 0^+2^{++}$
coherent $\pi$ exchange amplitudes. The $\pi$ exchange $\phi K^+ K^- n$ 
amplitudes were slowly varying and were either a non-resonant or broad 
resonance source. Thus the expected factor $\sim$ 100 suppression has 
become a factor 8 enhancement or factor $\sim$ 800 breaking of the OZI rule. 
All waves up to J=6 and L=4(114 waves) were searched for and except for one
S-wave and two D-waves which contributed to the three $2^{++}$ $\phi \phi$ 
resonances, nothing was found. The established resonance $f_4(2050)$ with 
$I^GJ^{PC} = 0^+4^{++}$ which could in principle be detected, other 
expected from the quark model states, and some reported but not established 
states were not detected. As stated above the t dependence showed that the 
three $2^{++}$ $\phi \phi$ resonances were produced by annihilation of the 
incident $\pi^-$ and a exchange $\pi^+$ to form the gluon bridge across 
the disconnection to the $\phi \phi$ final state. The annihilating pions 
contain only $u$ and $d$ quarks and the $\phi \phi$ is practically all $s$ 
and $\bar s$ quarks, except for the very small contribution due to the 
very small departure from ideal mixing. Thus it is a very clean flavor 
changing example. Fig.1 explains the powerful selectivity of the $\phi \phi$ 
system. Due to the $\phi$ spin, in addition to the Gottfried-Jackson and 
Trieman-Yang angles, the polar($\theta$) and azimuthal angles($\alpha$) of 
each $\phi$ due to its spin gave six significant angles and their correlations
 which yield the enormously selective signature for each wave, examples of 
which are shown.
\begin{figure}
\begin{center}               
\mbox{
   \epsfysize 4.4in
   \epsfbox{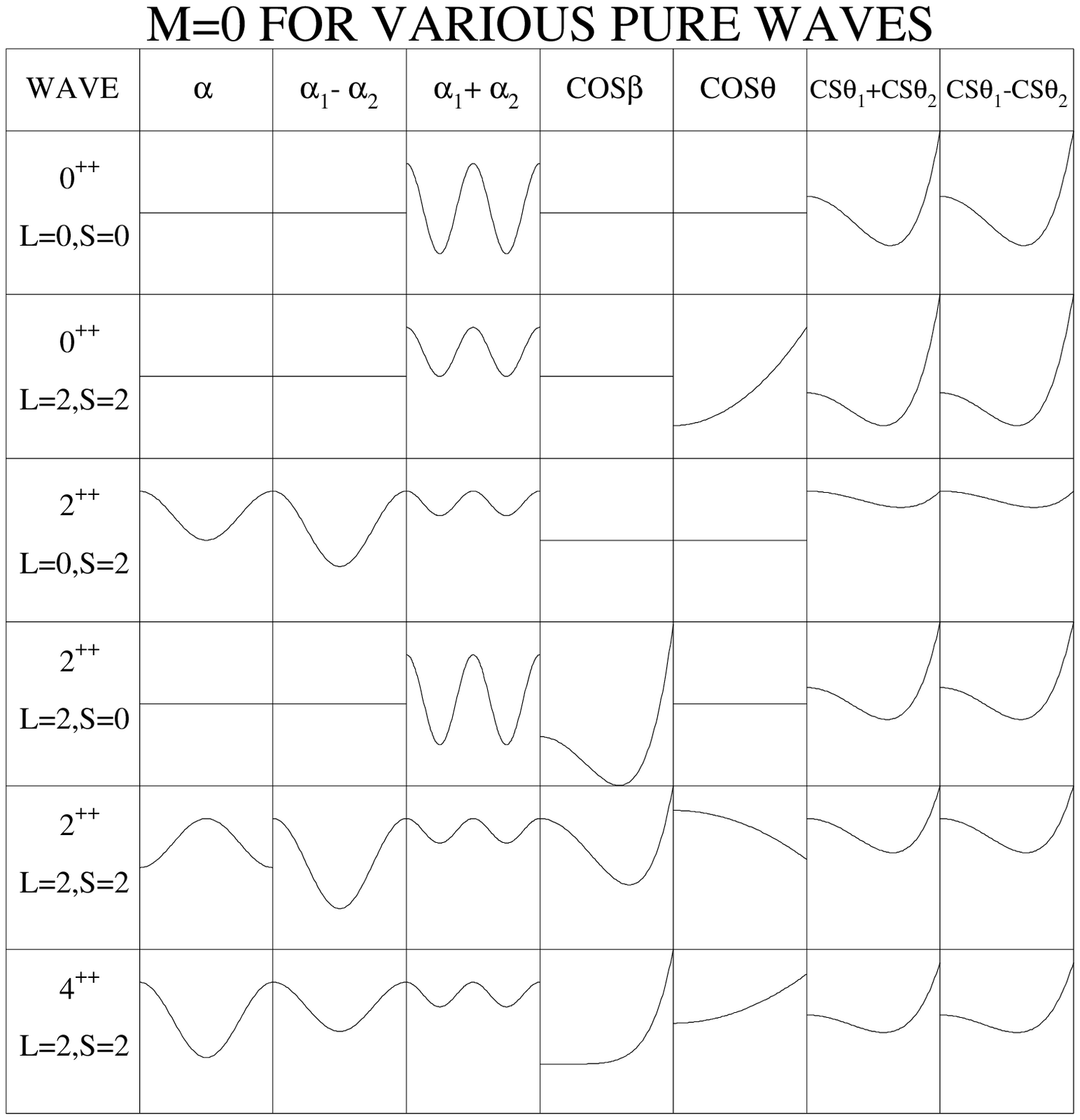}}
\end{center}
\caption{ Above we show 6 out of the 114 waves tried in our $\phi \phi$ partial
wave analysis. We show projections of the pure waves along: column 1 the 
$\phi$ azimuthal decay angle($\alpha$); column 2 the difference of azimuthal 
decay angles for $\phi_1$ and $\phi_2$; column 3 the sum; column 4 the 
cosine of the Gottfried Jackson angle($\beta$); column 5 the cosine of the 
polar $\phi$ decay angle($\theta$); column 6 the sum of the cosines of the 
polar decay angles of $\phi_1$ and $\phi_2$; column 7 the difference.}
\label{fig1}
\end{figure}

Reference \cite{phiphi4} concluded the OZI suppression was broken down by 
one primary $2^{++}$ glueball mixing with two quark states. The very strong 
glueball mixing effects on the quark states, which are not yet known, are
likely to change the character of these mixed states drastically. Hence 
attempts to identify these states via the quark model calculations appear
unjustified. The OZI rule argument has been attacked unsuccessfully by a 
number of theorists\cite{attack1,attack2}. However, to our knowledge, except 
for J = 0 where vacuum mixing could introduce flavor changing, the OZI rule 
works experimentally in all strong interaction cases where there is clean 
flavor changing annihilation between the initial and final states,
except where glue type resonances occur. It was also shown that by comparing
$K^-$ $p$ and $\pi^-$ $p$ interactions that the OZI rule was more or less
entirely broken down in $\pi^- p \rightarrow \phi \phi n$ \cite{Kbeam}
in the mass region investigated. It would require a flavor changing resonance 
to do this for J $>$ 0, and a glueball is the naturally expected choice. No 
other viable explanation of the data[2-4] has occurred since the initial 
results in 1982 \cite{phiphi2}. 

\section{Is The p p Central Production Related ?}

We now consider the evidence that a $4^{th}$ state may be related to the
above $\phi \phi$ experiment and analysis\cite{phiphi4}. Reference \cite{4pi} 
studied 4 $\pi$ channels in central $p$ $p$ interactions at 450 GeV/c. Their 
fig 1b shows a broad mass spectrum in the mass of $4 \pi^0$ which they consider
likely to be least affected by other effects. In a spin analysis they conclude
in fig 1e that the broad mass spectrum is $I^GJ^{PC} = 0^+2^{++}$ decaying
via a $f_2(1270) \pi \pi$ in a overall S-wave to $4\pi^0$. Fig 3f 
shows the same $I^GJ^{PC} = 0^+2^{++}$ $f_2(1270) \pi \pi$ wave in the 
$2\pi^+2\pi^-$ channel, which has the most statistics. They concluded that 
their analysis was consistent with one $f_2(1270) \pi \pi$  resonance with
a mass of 1950 MeV. It is for the $I^GJ^{PC} = 0^+2^{++}$ $f_2(1270) \pi \pi$ 
wave, that we would like to determine whether it could be related to previous 
$\phi \phi$ work\cite{phiphi4}. 

First we considered the possibility of fitting the $\phi \phi$ resonances
together with the $I^GJ^{PC} = 0^+2^{++}$ $f_2(1270) \pi \pi$ wave of 
Ref\cite{4pi}. The same K-matrix methodology used previously was again 
utilized\cite{phiphi4}. First we considered only three poles which was what 
was found sufficient in the $\phi \phi$ work. Our best fit was an unacceptable
9 $\sigma$ fit. One should keep in mind here that the physical mass spectrum 
of the $\phi \phi$ system was cut off at 2.038 GeV, so one could not 
experimentally measure below this value. However the lowest(2.010 GeV) 
$\phi \phi$ resonance width of 200 MeV extended below the threshold by 130 MeV.
However when we introduced a $4^{th}$ pole, we obtained a good 2$\sigma$ fit 
for the $\phi \phi$ and the new $f_2(1270) \pi \pi$ data\cite{4pi} combined. 
This new data has the same $I^GJ^{PC} = 0^+2^{++}$ quantum numbers as the 
$\phi \phi$ resonances, thus implying there is reasonable evidence suggesting 
they are related. In this fit the structure that appears around 2.4 GeV 
in both data sets connects the two systems(Fig.2-3). Requiring an additional
pole does not change the previous glueball interpretation presented in  
Ref\cite{phiphi4}. The number of glueballs plus glueball + quark mixed states 
cannot be determined in LGT without including quark loops. However having a 
$2^{++}$ glueball was the only viable explanation for the data published in 
1982\cite{phiphi2}. Nothing which has occurred in the two decades since 
contradicts this, but instead strengthens the case[3-4,10-12]. 

Now let us consider the details of the four pole fit. We have extracted the 
resonance parameters corresponding to each of the four poles which are 
displayed in Table 1.

\begin{center}
\begin{tabular}{|c|c|c|c|c|}\hline
\multicolumn{5}{|c|}{Table I - Resonance parameters}\\ \hline
Mass(GeV) & Width(GeV) &  $S_2$ &  $D_0$ &  $D_2$ \\ \hline
2.049$^{+.035}_{-.024}$   & .567$^{+.064}_{-.071}$ & 96\%$^{+4}_{-10}$ & 1\%$^{+7}_{-1}$ & 3\%$^{+9}_{-3}$ \\ \hline
2.123$^{+.015}_{-.033}$   & .294$^{+.056}_{-.055}$ & 89\%$^{+7}_{-11}$ & 2\%$^{+13}_{-2}$ & 9\%$^{+10}_{-8}$ \\ \hline
2.340$^{+.013}_{-.013}$   & .148$^{+.066}_{-.032}$ & 7\%$^{+15}_{-7}$ & 12\%$^{+14}_{-10}$ & 81\%$^{+3}_{-14}$ \\ \hline
2.412$^{+.028}_{-.032}$   & .362$^{+.100}_{-.053}$ & 45\%$^{+28}_{-7}$ & 21\%$^{+7}_{-19}$ & 34\%$^{+15}_{-20}$ \\ \hline
\end{tabular}
\end{center}

The first resonance best fit mass is barely above the lower boundry of the
$\phi \phi$ mass region, but due to the large width of the pole and its errors
it could easily be below. The other three resonances are clearly in the
$\phi \phi$ mass region. The $2^{nd}$ resonance seems similar to the previously
obtained $f_2(2010)$. It is approximately 89\% S-wave, but is shifted to higher
mass by about 100 MeV and is wider. The $3^{rd}$ resonance is similar to the
previously obtained $f_2(2300)$ shifted upward by about 40 MeV, but well
within the errors. The width is the same and it is about 95\% D-wave
consistent with the previous case. The $4^{th}$ resonance is similar to our 
previous $f_2(2340)$. However there are some differences in the division of 
the D-wave into the spin=0 and spin=2 states in the $3^{rd}$ and $4^{th}$ 
resonances. 

Fig.2a shows the $I^GJ^{PC} = 0^+2^{++}$ $f_2(1270) \pi \pi$ mass spectrum,
where the solid line is the 4 resonance combined fit given in Table I, and the 
dashed line is the 3 resonance fit. The dotted curve will be explained in
the next section. Fig.2b shows the $I^GJ^{PC} = 0^+2^{++}$ spin 2 S-wave 
$\phi\phi$ mass distribution. Again the solid line is the 4 resonance fit 
given in Table I and the dashed line is the 3 resonance fit. This is true 
for all Figs.2-3. Fig.2c shows the $I^GJ^{PC} = 0^+2^{++}$ spin 0 D-wave 
$\phi\phi$ mass spectrum. Fig.2d shows the $I^GJ^{PC} = 0^+2^{++}$ spin 2 
D-wave $\phi\phi$ mass spectrum. 

The S-wave was used as a phase reference. Fig.3a shows the $2^{++}$ spin 0 
D-wave phase relative to the S-wave as a function of $\phi\phi$ mass. Fig.3b
shows the $2^{++}$ spin 2 D-wave phase. The three resonance fit gives a very 
crude qualitative agreement, but must be discarded as it fails quantitatively 
by 9 $\sigma$, whereas the four resonance fit is a 2 $\sigma$ fit.

Fig.3c shows the Argand diagram for the $I^GJ^{PC} = 0^+2^{++}$ $f_2(1270) 
\pi \pi$ amplitude. The Argand diagrams for the three $I^GJ^{PC} = 0^+2^{++} 
\phi \phi$ amplitudes are shown in Fig.3d. These Argand diagrams show clear
resonance behavior. The scale on the Argand plots is the square root of the
scales of Fig. 2. The fact that the four resonance fit is a good fit
which ties together the structure in the $f_2(1270) \pi \pi$ and the 
$\phi \phi$ structure near 2.4 GeV, while retaining the basic characteristics 
of the orginal three $\phi \phi$ resonance states implies it is likely they 
are related, and due to the same glueball and mixing mechanism which explains
the $\phi \phi$ states.

\begin{figure}
\begin{center}               
\mbox{
   \epsfysize 4.4in
   \epsfbox{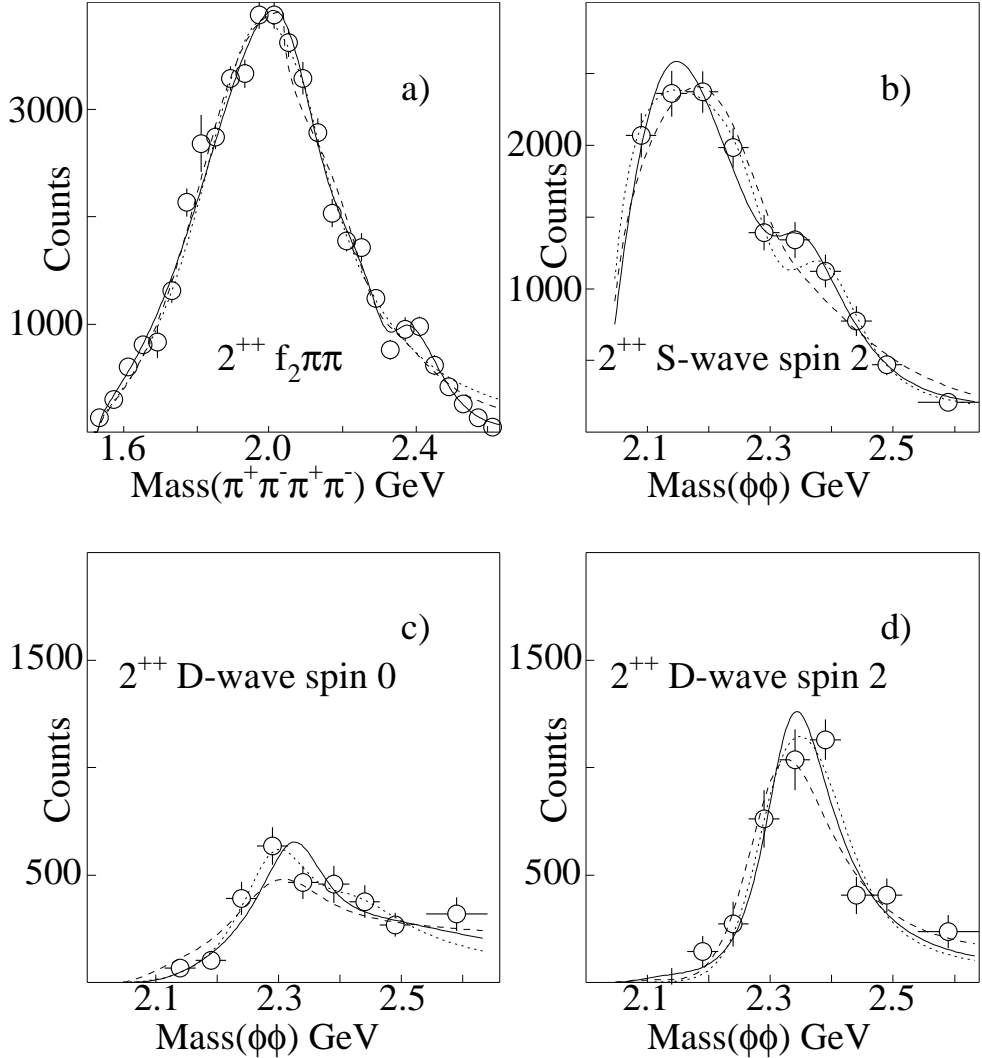}}
\end{center}
\caption{The $I^GJ^{PC} = 0^+2^{++}$ mass spectra of: a) Mass(
$\pi^+\pi^-\pi^+\pi^-$) for  $2^{++}f_2(1270) \pi \pi$; b) Mass($\phi \phi$)
for $2^{++}$ S-wave spin 2; c) Mass($\phi \phi$) for $2^{++}$ D-wave spin 0
; d) Mass($\phi \phi$) for $2^{++}$ D-wave spin 2. The solid curve is the 
four resonance fit(table I); dashed curve is a three resonance fit; and 
dotted curve is a separate three resonance fit to the three $\phi\phi$ 
waves, and a separate fit with one resonance to the $f_2(1270) \pi \pi$ 
wave. Note this separate fit for the $f_2(1270) \pi \pi$ is particularly 
poor in the region of 2.3 to 2.6 GeV, whereas the 4 pole(resonance) combined 
fit(solid line) is a 2 $\sigma$ fit(see text)} 
\label{fig2}
\end{figure}

\begin{figure}
\begin{center}               
\mbox{
   \epsfysize 4.4in
   \epsfbox{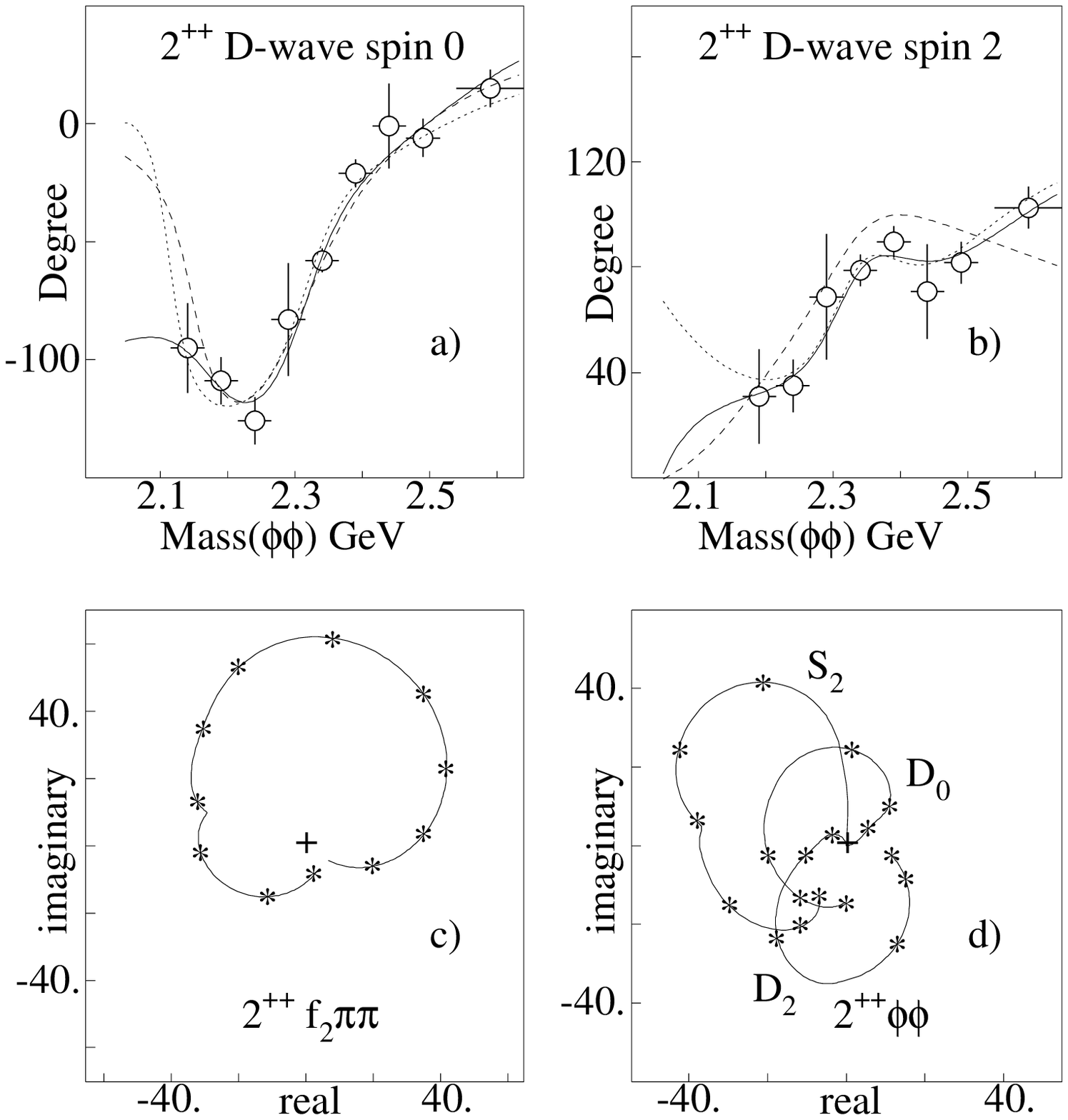}}
\end{center}
\caption{a) and b) Show the phase behavior of the D-wave $\phi \phi$ amplitudes
measured with respect to spin2 S-wave for the three fits described in Figure
2 caption and text: a) $2^{++}$ spin 0 D-wave; b) $2^{++}$ spin 2 D-wave.
c) and d) Show Argand amplitudes which are generated by the four 
pole(resonance) fit. Each * is at a 0.1 GeV marking starting with 1.5 GeV for
c) the $2^{++} f_2 \pi \pi$ Argand amplitude, while d) markings start at 2.1
GeV for the three $\phi \phi$ waves spin2 S-wave($S_2$), spin0 D-wave($D_0$)
, and spin2 D-wave($D_2$).} 
\label{fig3}
\end{figure}

\section{Suppose they are not related}

We can obtain a four resonance fit and have no relationship between the
resonance in $f_2(1270) \pi \pi$ and the three $\phi \phi$ resonances.
We show in Fig.2a-3b as a dotted line the four resonances in which there is
no connection between $f_2(1270) \pi \pi$ and $\phi \phi$. The $\phi \phi$ n
data contains additional statistics obtained beyond \cite{phiphi4} and a
new analysis which is consistent \cite{phiphi4}. This fit is a 6 $\sigma$ fit,
where the most of the $\sigma$'s arise from Ref\cite{4pi} not fitting the
structure in their data near 2.4 GeV which overlapped with the $\phi \phi$ mass
 region. One could add a second resonance in the $f_2(1270) \pi \pi$  or
attribute that to some other effect. However even if one assumes that the 
$f_2(1270) \pi \pi$ is not related to the $\phi \phi$ that in no way 
contradicts the conclusions drawn on the basis of the $\phi \phi$ resonances
data. Thus in the two decades of physics work following the initial strong 
evidence explainable only with the effects caused by a $2^{++}$ glueball, all 
that occurred both in extension of that work, and all other relevant works 
both experimental and theoretical in the region of 2 GeV mass, has strengthened
the initial evidence. Nothing published has successfully contradicted it, or 
produced a viable alternative explanation of the $\phi \phi$ data.

A much quoted classic paper on the LGT glueball mass spectrum by Morningstar 
and Peardon\cite{LGT4} using SU(3) pure gauge theory finds that there can be 
glueball states consisting of one to several glueballs degenerate in mass.
However their investigations conclude the multi-glueball states would be high
enough in mass so that the predicted lowest lying $I^GJ^{PC} = 0^+2^{++}$
glueball would be a single glueball which could mix with quark-built states
in the $\sim$2 GeV mass region. The mass they determined for the $2^{++}$ 
glueball is $2.40\pm{.15}$ GeV which is consistent with the  $\phi \phi$ 
resonance data.

A recent discussion on the $\phi \phi$ resonance states with Michael Creutz 
led to a private communication he sent us, from which we quote since it should
be of interest to the readers. ``I am glad to hear that you are continuing to
pursue the $\phi \phi$ states as glueball candidates. I personally believe 
that these states are indeed likely to contain substantial glue, although I
expect there will be some mixing with quark states as well. This mixing will
make a complete analysis very complicated, but the glue component should
exist and will appear in all the states that are mixed''\cite{Creutz}.

\section{Conclusion}

The breakdown of the OZI suppression by a factor $\sim$800 is enormous. It is
strikingly extremely selective since only the three $\phi \phi$ resonances
all of which have $I^GJ^{PC} = 0^+2^{++}$, the quantum numbers of the $2^{++}$
glueball, and nothing else is found in the analysis up to J=6 and L=4 
(114 waves) in a mass region consistent with the predicted $2^{++}$ glueball
mass. This has only been explained by a glue resonance($2^{++}$ glueball).
No other viable alternative has been published in two decades since the initial
results in 1982\cite{phiphi2} which have been strengthened with time[1-4]and
[10-13]. The 450 GeV/c pp central production results of \cite{4pi} are
explainable by a simple extension of the previous $\phi \phi$ resonance
analysis\cite{phiphi4} to include an additional $4^{th}$ pole which results
in a 2 $\sigma$ fit for the combined data.

Acknowledgements

This research was supported by the U.S. Department of Energy under Contract 
No. DE-AC02-98CH1088 and the City College of New York Physics Department.
Except for minor changes it is reported in BNL-72371-2004 Formal Report by
R.S. Longacre and S.J. Lindenbaum on the US DOE web June 2004.

\end{document}